\documentclass[a4paper]{jpconf}
\usepackage{graphicx,amsmath,amssymb}
\begin{document}
\def\vac{|{\rm vac}\rangle}
\title{Charge instabilities of the two-dimensional Hubbard model with attractive nearest neighbour interaction}

\author{Raymond Fr\'esard}
\address{Laboratoire CRISMAT, UMR 6508 CNRS-ENSICAEN, Caen, France}
\author{Kevin Steffen and Thilo Kopp}
\address{Center for Electronic Correlations and Magnetism, EP VI, Institute of Physics, University of Augsburg, 86135 Augsburg, Germany}

\ead{Raymond.Fresard@ensicaen.fr}

\begin{abstract}
Attractive non-local interactions jointly with repulsive local interaction in
a microscopic modelling of electronic Fermi liquids generate a competition
between an enhancement of the static charge susceptibility---ultimately 
signalling  charge instability and phase separation---and its correlation
induced suppression. We analyse this scenario through the investigation of the
extended Hubbard model on a two-dimensional square lattice, using the spin
rotation invariant slave-boson representation of Kotliar and Ruckenstein. 
The quasiparticle density of states, the renormalised effective mass and the
Landau parameter $F_0^s$ are presented, whereby the positivity of  $F_0^s-1$
constitutes a criterion for stability. Van Hove singularities in the density
of states support possible charge instabilities. A (negative) next-nearest
neighbour hopping parameter $t'$ shifts their positions and produces a tendency
towards charge instability even for low filling whereas the $t'$-controlled
particle-hole asymmetry of the correlation driven effective mass is small. 
A region of instability on account of the attractive interaction $V$ is
identified, either at half filling in the absence of strong electronic
correlations or, in the case of large on-site interaction $U$, at densities
far from half filling.

\end{abstract}

\section{Introduction}

Attractive nearest neighbour interactions are commonly introduced in electronic
lattice models in order to study (unconventional) superconductivity. However,
an attractive  non-local interaction can also cause a very different
phenomenon in the electronic system; in particular, it can generate charge
instabilities. In this paper we focus on the tendency towards charge
instabilities in two-dimensional (2D) electronic systems which are
simultaneously characterised by electronic correlations. Through their
correlations the electrons constitute a Fermi liquid, with non-zero Landau
parameters and enhanced effective mass. The Fermi liquid behaviour is expected
to break down, especially if the correlations are strong close to half filling
or when the attractive nearest neighbour interaction dominates. 

It is quite conceivable that the on-site interaction of electrons is repulsive
whereas the nearest neighbour interaction becomes attractive. Typically, the
on-site Coulomb interaction at transition-metal ions is strongly screened by
the polarisation of neighbouring atoms, e.g., oxygen ions in
cuprates~\cite{Boer84,Brink97}. Yet the screening depends on the local
excitations of a cluster of atoms---for example, on the virtual excitations
related to charge transfer in CuO$_6$-octahedra---and can be different for
nearest neighbour Coulomb interaction as compared to the local screening or the
Lindhard-type screening for longer distances. Coupling to other degrees of
freedom~\cite{Mic90}, such as a strong electron-lattice coupling, may in
combination with screening produce an attractive nearest neighbour interaction,
as in certain iron-pnictides~\cite{Saw09}. 

In recent years, metallic two-dimensional electron systems were identified at
the interface between two insulating films in oxide
heterostructures~\cite{Oht04,Thi06}. In contrast to semiconductor interfaces
the electronic states in the oxide materials are confined to few atomic layers
in the vicinity of the interface (see, e.g. Ref.~\cite{Sin09}). Electronic
correlation effects have been observed in scanning tunnelling
spectroscopy~\cite{Bre10}, and the electron system was characterised as a 2D
electron liquid. Moreover, the compressibility, which is manifestly related to
the Landau Fermi-liquid parameter $F_0^s$, was experimentally observed to be
negative in a regime of low charge carrier density~\cite{Tin12}---the latter
being tuned by a backgate bias. 

For the Hubbard model~\cite{Hub63}, Landau parameters were calculated by
Vollhardt~\cite{Vol84} within Gutzwiller approximation. Recently,
Lhoutellier \textit{et al.}~\cite{Lhou15} calculated the Landau parameters
$F_0^s$ and $F_0^a$ for an extended three-dimensional Hubbard model within a
spin rotation invariant generalisation~\cite{Li89,FW} of the Kotliar and
Ruckenstein slave-boson (KRSB) representation~\cite{Kot86}. For a 2D system
the Hubbard model extended by intersite Coulomb interaction and
electron-phonon coupling was shown, within a KRSB evaluation, to  produce
inhomogeneous polaronic states~\cite{Pav06}, a scenario which may well be
realised in some heterostructures. In the present paper we are concerned with
a 2D system and we use the same scheme as in Ref.~\cite{Lhou15} to evaluate
the effective mass $m^*$ and $F_0^s$ with focus on attractive non-local
interactions and the ensuing charge instabilities. Specifically, we
investigate their respective filling dependence. Do electronic correlation
effects compete with the tendency towards charge separation? 

The KRSB representation was introduced to realise the interaction driven
Brinkman-Rice metal-to-insulator transition \cite{Bri70} but since then, it
has been successfully used to analyse and characterise antiferromagnetic
\cite{Lil90,Kot00}, ferromagnetic \cite{Doll2}, spiral
\cite{Fre91,Arr91,Fre92,Igoshev13}  and striped \cite{SeiSi,Sei02,Rac06,RaEPL} 
phases. KRSB evaluations have been tested against quantum Monte Carlo
simulations: A quantitative agreement for charge structure factors was
demonstrated \cite{Zim97} and, for example, a very good agreement on the
location of the metal-to-insulator transition for the honeycomb lattice has been
shown \cite{Doll3}. Also the comparison of ground state energies 
to numerical solutions~\cite{Fre91} or exact diagonalisation data are
excellent~\cite{Fre92}. 

The paper is organised as follows: The extended Hubbard model is introduced in
Sec.~\ref{sec:hub}, together with its Kotliar and Ruckenstein spin rotation
invariant slave-boson representation. In Sec.~\ref{sec:saddle} the
saddle-point approximation is presented, jointly with the resulting system 
of coupled nonlinear equations. Fluctuations are captured within the
one-loop approximation, Sec.~\ref{sec:loop}, which allows to determine
analytically the Landau parameter $F_0^s$ at half filling for the pure Hubbard
model. Numerical results are discussed in Sec.~\ref{sec:resu}, 
where we address the filling dependence of $F_0^s$ and charge instabilities. 
The paper is summarised in Sec.~\ref{sec:summa}.

\section{Extended Hubbard model}
\label{sec:hub}

Anderson's suggestion that the simplest model for the $d$-electrons within the
CuO$_2$ layers common to high-T$_c$ superconductors is the Hubbard model
\cite{And87} stimulated tremendous research on its properties. However, in the
Hubbard model the Coulomb interaction is restricted to the on-site contribution
only. In fact, Hubbard himself already argued \cite{Hub63} that for transition metals the matrix elements corresponding to nearest neighbour Coulomb repulsion are relatively large and cannot be disregarded \textit{a priori}. The extended Hubbard model,
\begin{equation}
H = \sum_{i,j,\sigma}t_{ij}^{\phantom{\dagger}}c_{i\sigma}^{\dagger}c_{j\sigma}^{\phantom{\dagger}} +U\sum_{i}\left(n_{i\uparrow}^{\phantom{\dagger}}-\frac12 \right) \left(n_{i\downarrow}^{\phantom{\dagger}} -\frac12 \right) +\frac12 \sum_{i,j}V_{ij}^{\phantom{\dagger}}(1-n_{i}^{\phantom{\dagger}}) (1-n_{j}^{\phantom{\dagger}}),
\label{eq:model}
\end{equation}
takes this notion into account by including intersite Coulomb $V_{ij}$ interactions. Although these elements decay fast with increasing distance $|\vec{R}_i-\vec{R}_j|$, they extend in general beyond nearest neighbours. Here $c_{i\sigma}^\dagger$ denotes electron creation operators at site $i$ with spin $\sigma$, and $n_{i\sigma}^{\phantom{\dagger}}=c_{i\sigma}^{\dagger}c_{i\sigma}^{\phantom{\dagger}}$. The particle-hole symmetric form for both density-density interaction terms is used throughout this work.

In our numerical evaluations below, we restrict the matrix elements $t_{ij}$ to $-t$ for $(i,j)$ a pair of nearest neighbour sites and to $-t'$
for next-nearest neighbour pairs on a square lattice. All other $t_{ij}$ are set to zero.
While the bare intersite Coulomb interactions are repulsive, the effective parameters $\{V_{ij}\}$ may become attractive if the coupling to other subsystems is considered~\cite{Mic90}. An example is strong electron-lattice coupling, which induces a local lattice deformation surrounding the electron. Under certain conditions such electronic polarons may attract each other as, for example, in some iron-pnictides \cite{Saw09}. Since we are interested in instabilities indicated by $F_0^s\leq -1$, we limit our numerical calculations in this paper to $V_{ij}\le 0$. 

In the spin rotation invariant (SRI)  Kotliar and Ruckenstein slave-boson (KRSB) representation~\cite{Li89,FW}, which we adopt for our study, one introduces the auxiliary canonical fermionic $f_{\sigma}$, and bosonic $e$, $p_{0}$, ${\vec p}$, and $d$ particles to represent the physical states as: 
\begin{eqnarray}
|0\rangle&=& e^{\dagger}\vac \nonumber\\
|\sigma\rangle&=&\sum_{\sigma'}p^{\dagger}_{\sigma\sigma'}
f^{\dagger}_{\sigma'} |\mbox{vac}\rangle {\;\;\;\;\;\;\;\;\;\;\sigma
=\uparrow,\downarrow} \nonumber\\
|2\rangle&=& d^{\dagger}
f^{\dagger}_{\uparrow} f^{\dagger}_{\downarrow}\vac \,,
\end{eqnarray}
with $\underline{p^{\dagger}}=\frac12\sum_{\mu=0}^3
p^{\dagger}_{\mu}\underline{\tau}^{\mu}$, and $\underline{\tau}^{\mu}$ the
Pauli matrices. In terms of these auxiliary
 operators  the Hamiltonian Eq.~(\ref{eq:model}) reads
\begin{eqnarray}
H &=& \sum_{i,j}t_{ij} \sum_{\sigma\sigma'\sigma_{1}}z^{\dagger}_{i\sigma_{1}\sigma}f^{\dagger}_{i\sigma}f^{\phantom{\dagger}}_{j\sigma'}z^{\phantom{\dagger}}_{j\sigma'\sigma_{1}} +U\sum_{i}\left(d_{i}^{\dagger}d^{\phantom{\dagger}}_{i} -\frac12 \sum_{\sigma} f^{\dagger}_{i\sigma}f^{\phantom{\dagger}}_{i\sigma} +\frac{1}{4} \right)\nonumber\\
&+& \frac{1}{4} \sum_{i,j} V^{\phantom{\dagger}}_{ij}\!\left[\!\left( 1-\sum_{\sigma} f^{\dagger}_{i\sigma}f^{}_{i\sigma}\!\right)\!Y_j+Y_i\!\left( 1- \sum_{\sigma}f^{\dagger}_{j\sigma}f^{\phantom{\dagger}}_{j\sigma} \!\right)\! \right] \,.
\label{eq:sbmodel}
\end{eqnarray}
The spin and charge degrees of freedom are mapped onto bosons in this representation; further details are given in Refs.~\cite{FKW} 
and~\cite{Lhou15}. We now suppress the site indices in expressions where their reintroduction is self-evident. Above, we introduced the hole doping operator
\begin{equation}
Y \equiv e^{\dagger} e^{\phantom{\dagger}} -d^{\dagger} d^{\phantom{\dagger}}\,,
\end{equation}
while $\underline{z}$ is given by:
\begin{equation}\label{eqrf:forz}
\underline{z} \equiv e^\dagger\underline{L}^{\phantom{\dagger}} M^{\phantom{\dagger}}\underline{R}^{\phantom{\dagger}} \underline{p}^{\phantom{\dagger}} +\underline{\tilde{p}}^{\dagger}\underline{L}^{\phantom{\dagger}}M^{\phantom{\dagger}} \underline{R}^{\phantom{\dagger}}  d 
\end{equation}
where 
$\tilde{p}^{\phantom{\dagger}}_{\sigma\sigma'}=(\delta_{\sigma,\sigma'} - 
\delta_{\sigma,-\sigma'})p^{\phantom{\dagger}}_{-\sigma',-\sigma}$ and
\begin{eqnarray} \label{eqrf:z_SRI}
M &=& \left[1+ e^\dagger e + \sum_{\mu} p^\dagger_{\mu}p^{\phantom{\dagger}}_{\mu} + d^\dagger d \right]^{\frac12}, \nonumber\\
\underline{L}^{\phantom{\dagger}} &=& \left[\left(1- d^\dagger d\right) \underline{1}-2 \underline{p}^{\dagger}\underline{p}^{\phantom{\dagger}}\right]^{-\frac12}, \nonumber\\
\underline{R} &=& \left[\left(1- e^\dagger e\right) \underline{1}-2 \underline{\tilde{p}}^{\dagger}\underline{\tilde{p}}^{\phantom{\dagger}}\right]^{-\frac12} .
\end{eqnarray}

The tensor $\underline{z}$ may also be expanded in terms of Pauli matrices as 
$\underline{z}=\sum_{\mu=0}^3 z_{\mu}\underline{\tau}^{\mu}$.
The new Fock space depends in this representation on eight auxiliary
operators. The resulting unphysical degrees of freedom may be eliminated
through five local constraints,
\begin{eqnarray}
\label{eqrf:Qcompl}
e^{\dagger}e+\sum_{\mu} p^{\dagger}_{\mu}p^{\phantom{\dagger}}_{\mu}+ d^{\dagger}d &=& 1\,, \\
\label{eqrf:Qscal}
\sum_{\sigma} f^{\dagger}_{\sigma} f^{\phantom{\dagger}}_{\sigma} &=&\sum_{\mu} p^{\dagger}_{\mu}p^{\phantom{\dagger}}_{\mu}+ 2 d^{\dagger}d\,, \\
\label{eqrf:Qvect}
\sum_{\sigma,\sigma'} f^{\dagger}_{\sigma'} {\vec \tau}_{\sigma\sigma'}f^{\phantom{\dagger}}_{\sigma} &=& p^{\dagger}_{0} {\vec p}+ {\vec p}^{\,\dagger} p^{\phantom{\dagger}}_{0}-i {\vec p}^{\,\dagger} \times {\vec p}^{\phantom{\dagger}}\,,
\end{eqnarray}
which are enforced within the path integral formalism and have to be satisfied for each lattice site.

In the SRI KRSB representation the phases of the $e$ and $p_{\mu}$ bosons are
gauged away by promoting all constraint parameters to fields \cite{FW}, and
they remain as radial slave-boson fields~\cite{Fre01}. Their exact
expectation values are generically non-vanishing even though bose condensation
is excluded by local gauge invariance~\cite{Kop07}. In contrast, the slave-boson field corresponding to double occupancy $d$ is still
complex~\cite{FW,Jol91,Kot92}. 

The saddle-point approximation introduced in the next section is exact in the
large degeneracy limit, with Gaussian fluctuations generating the $1/N$
corrections~\cite{FW}, and obeys a variational principle in the limit of large
spatial dimensions. In this limit the Gutzwiller approximation becomes exact
for the Gutzwiller wave function and longer ranged interactions are static and
reduce to their Hartree approximation~\cite{Met89,Mul89}. On these grounds,
the approximation used here to the extended Hubbard model Eq.~(\ref{eq:model})
complies with a variational principle in the limit of large spatial
dimensions. 

Due to the formal properties mentioned above, our approach covers properties
of strongly correlated electrons, as, e.g., the suppression of the
quasiparticle residue and the formation of a charge gap at the Mott-Hubbard
transition, and the Brinkman-Rice transition~\cite{Bri70} to an insulating
state at half filling with increasing on-site Coulomb interaction. The
impact of the non-local interaction on the latter transition is of
particular interest. 
 
\section{Saddle-point approximation}
\label{sec:saddle}

\begin{figure}[t]
\begin{center}
\includegraphics[width=34pc]{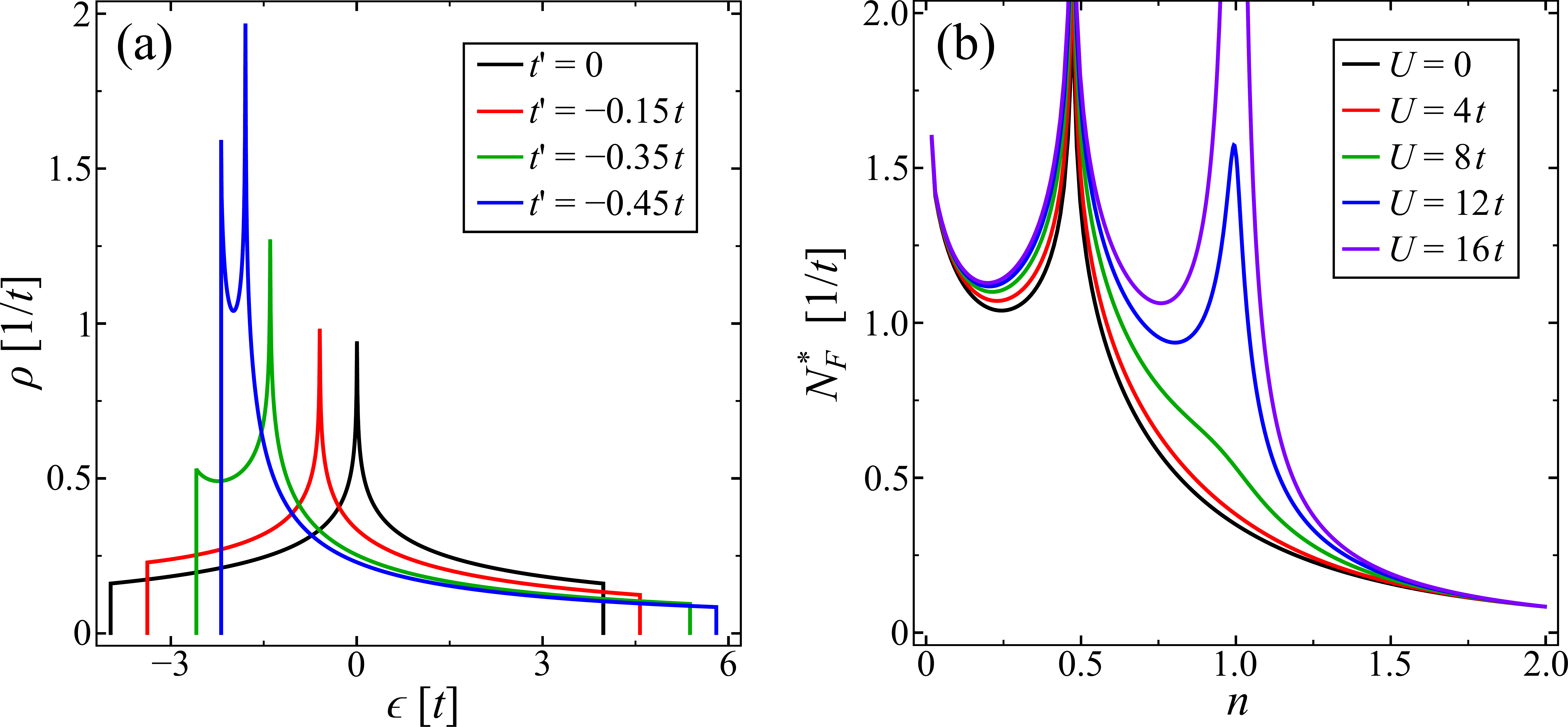}
\end{center}
\caption{\label{fig:dos}Density of states (DOS). \textit{a}) DOS $\rho(\epsilon)$
of the square lattice  for various values of $t'$ as a function of band energy $\epsilon$.
\textit{b}) Effective DOS at the Fermi edge $N^*_F$ for $t'=-0.45\,t$ and varying on-site Coulomb interactions $U$
as a function of filling $n$. The effective DOS is independent of the nearest neighbour interaction $V$.}
 \end{figure}

Ideally the functional integrals should be calculated exactly. Regarding spin
models this has been achieved for the Ising chain \cite{Fre01}, but in the
case of interacting electron models exact evaluations could be performed on
small clusters only, either using the Barnes representation \cite{Kop07}, or the
Kotliar and Ruckenstein representation \cite{Kop12}. Such a calculation remains
challenging on lattices of higher dimensionality, and we here rather resort to the saddle-point
approximation. The presentation here for the 2D electronic system follows closely that 
of Ref.~\cite{Lhou15} where a 3D extended Hubbard model was investigated. 
In the translational invariant paramagnetic phase all
the local quantities are site independent, and the action at
saddle-point reads ($\beta=1/k_BT$),
\begin{equation}\label{eq:spact}
S = \beta L \left(S_B + S_F + \frac{1}{4} U\right)\,,
\end{equation}
where $L$ is the number of lattice sites and
\begin{eqnarray}
\label{sb}
S_{B}&=&\alpha(e^{2}+d^{2}+p_{0}^{2}-1)-\beta_{0}(p_{0}^{2}+2d^{2})
+Ud^{2}+ \frac12 V_{0}Y\,, \\
\label{sf}
S_{F}&=&-\frac{1}{\beta}\displaystyle
\sum_{\vec k,\sigma}\ln\left(1+e^{-\beta E_{{\vec k}\sigma}}\right)\,.
\end{eqnarray}
Here $\alpha$ and $\beta_0$ are site-independent Lagrange multipliers that
enforce the constraints Eq.~(\ref{eqrf:Qcompl}) and Eq.~(\ref{eqrf:Qscal}),
respectively.  
For the extended Hubbard model (\ref{eq:model}) the quasiparticle
dispersion in Eq.~(\ref{sf}) reads:
\begin{equation}\label{eqrf:dispp}
E_{{\vec k}\sigma}=z_{0}^{2}t_{\vec k}+\beta_{0}-\frac12 U
-\frac12 V_{0} Y -\mu\;\; ,
\end{equation}
Here $z_0^2$ represents the inverse effective mass $m/m^*$ correction, 
exclusively caused by correlation effects.
The Fourier transform of the intersite Coulomb 
repulsion is
\begin{equation}
\label{eq:V0}
V_{\vec{k}} = \frac{1}{L} \sum_{i,j} V_{ij} e^{-i \vec{k}\cdot
(\vec{R}_j-\vec{R}_i)} \, .
\end{equation}
It is worth mentioning that only $V_{\vec{k}=0}$ enters Eq.~(\ref{eqrf:dispp}).
The saddle-point equations following from Eq.~(\ref{eq:spact}) read:
\begin{eqnarray}
\label{eq:speqss}
p_{0}^{2}+e^{2}+d^{2}-1&=&0, \nonumber\\
p_{0}^{2}+2d^{2}&=&n, \nonumber\\
\frac{1}{2e}\frac{\partial z_{0}^{2}}{\partial
   e}\bar{\varepsilon}+\frac12 V_{0}\,(1-n)\frac{1}{2e}
   \frac{\partial Y}{\partial e}&=&-\alpha, \\
\frac{1}{2p_{0}}\frac{\partial z_{0}^{2}}{\partial p_{0}}
\bar{\varepsilon}&=& \beta_{0}-\alpha, \nonumber\\
\frac{1}{2d}\frac{\partial z_{0}^{2}}{\partial d}\bar{\varepsilon}
+\frac12 V_{0}\,(1-n)\frac{1}{2d}\frac{\partial Y}{\partial d}&=&
2 \beta_{0}-\alpha-U. \nonumber
\end{eqnarray}
Here we have introduced the averaged kinetic energy,
\begin{equation}
\label{eq:varepsilon}
\bar{\varepsilon}=\int d\epsilon \rho (\epsilon)\epsilon
f_{F}(z_{0}^{2}\epsilon +\beta_{0}-\frac12 U
-\frac12 V_{0} Y -\mu)\,,
\end{equation}
the determination of which involves the density of states $\rho(\epsilon)$ and
$f_F(\dots)$ is the Fermi function. The density of states (DOS) is displayed in Fig.~\ref{fig:dos}(a)
for various values of $t'$.
With $y \equiv (e+d)^2$ and the doping away from half filling
$\delta=1-n$, the equations Eqs.~(\ref{eq:speqss}) may be merged into a single one:
\begin{equation}\label{eq:spfinal}
y^{3}+(u-1)y^{2}=u\delta^{2}\,,
\end{equation}
where the Coulomb parameter is conveniently expressed in units of $U_{0}$ as
$u=U/U_{0}$ with
\begin{equation}
U_{0}=-\frac{8}{1-\delta^{2}}\,\bar{\varepsilon}\,.
\end{equation}
The three solutions of Eq.~(\ref{eq:spfinal}) have been discussed in much
detail at zero temperature in Ref.~\cite{FW}, with the result that they entail
one single critical point located at the Brinkman-Rice transition
\cite{Bri70}. It signals a second order transition. At finite temperature, a
line of first order  transitions with coexistence of phases is found
\cite{Doll3}. On the contrary, a finite Hund's coupling in a
two-band model induces coexistence of one good and one poor metallic phase
at and in the vicinity of half filling at zero temperature \cite{Lamb}. 

Returning to the one-band model at $\delta=0$, the inverse effective mass
$m/m^* = z_0^2$ may be obtained as 
\begin{equation}
z_0^2 = 1-u^2\,.
\end{equation}
Therefore a metal-to-insulator transition occurs at half filling for a given
lattice at $U_{c}$, which is defined as follows
\begin{equation}
U_c\equiv\lim_{\delta \rightarrow 0}U_{0}=-8\bar{\varepsilon}\,.
\label{uc}
\end{equation}

A remarkable property of the paramagnetic phase we consider is that  $V_{ij}$
elements do not enter explicitly
Eq.~(\ref{eq:spfinal}).
Hence, in this paramagnetic phase, the intersite
interactions only influence the fluctuations and do not change electron
localisation due to strong on-site interaction $U$. In particular, neither a
nearest neighbour Coulomb interaction $V_{<i,j>}$  nor a long-ranged one
$V(i-j)$ has influence on the Mott gap as discussed by Lavagna
\cite{Lav90}. Furthermore, the double 
occupancy is in exact agreement with the Gutzwiller approximation as
derived by Vollhardt, W\"olfle and Anderson \cite{Vol87}. In the present
case of a 2D square lattice the double occupancy vanishes at half filling
for $U_c=12.96\,t$ with $t'=0$, $U_c=13.06\,t$ with $t'=-0.15\,t $, $U_c=13.4\,t$ with
$t'=-0.35\,t$, and for $U_c=13.62\,t$ with $t'=-0.45\,t$. Thus, the location of the
Brinkman-Rice transition \cite{Bri70} shows little dependence on
$t'$. Somewhat surprisingly, the effect of the $t'$-induced gradual depression
of the DOS at half filling 
and, consequently, the depression of $U_c$ is opposite to the one found for the semi-metallic
honeycomb lattice, for which $U_c=12.6\,t$ was obtained \cite{Doll3}.

We here emphasise that Eq.~(\ref{eq:spfinal}) is not
  only insensitive to $V_{ij}$ elements, but also to the representation of the
  Hubbard interaction. Indeed, while it is commonly directly expressed in
  terms of double occupancy, the presently used particle-hole symmetric form
  results in the same equation. Hence, the energy depends on the precise
  form of the interactions but the slave-boson saddle-point values do not. 

\begin{figure}[t]
\begin{minipage}{18pc}
\hskip-0.2cm\includegraphics[width=18pc]{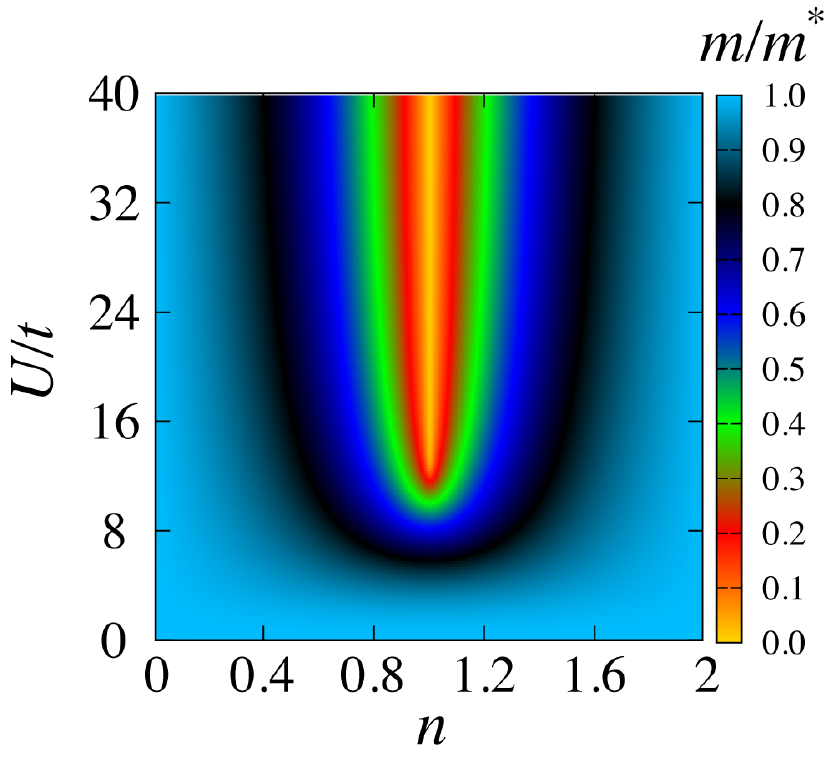}
\caption{\label{fig:z00}Inverse mass renormalisation for the square lattice with $t'=0$
as a function of $n$ and $U$. 
The effective mass $m^*$ is independent of the nearest neighbour interaction $V$.}
\end{minipage}\hspace{2pc}
\begin{minipage}{18pc}
\hskip-0.2cm\includegraphics[width=18pc]{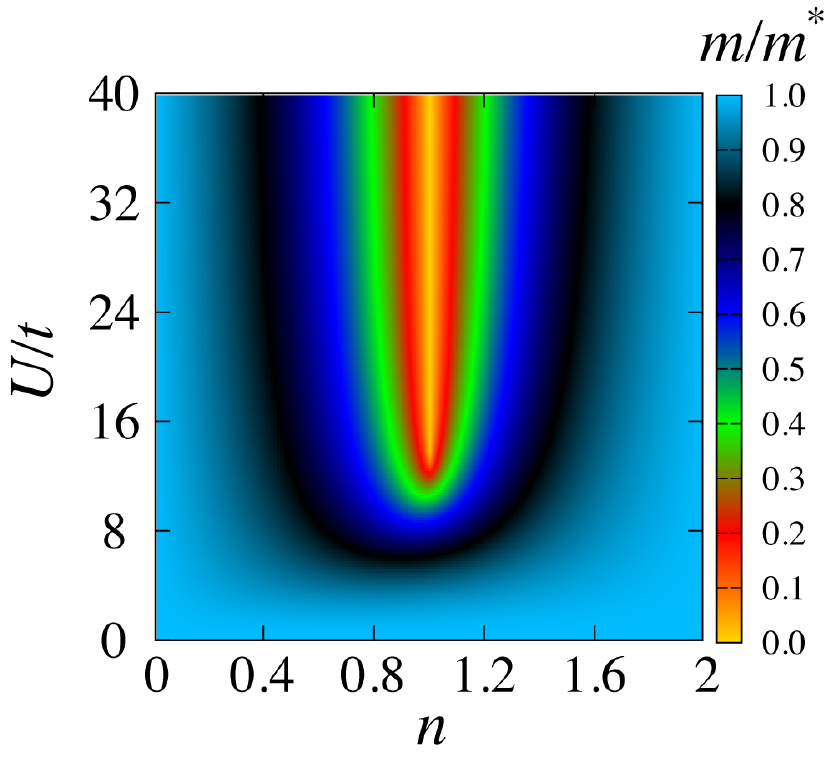}
\caption{\label{fig:z0p45}Inverse mass renormalisation for the square lattice with $t'=-0.45\,t$. 
The effective mass $m^*$ is independent of the nearest neighbour interaction $V$.}
\end{minipage} 
\end{figure}

\section{One-loop approximation to the charge response function}
\label{sec:loop}

The mapping of all degrees of freedom onto bosons allows for directly
evaluating the charge response function. According to, e.~g.,
Ref.~\cite{Lhou15}, the density fluctuations may be expressed as 
\begin{equation}
\delta N \equiv  \sum_{\sigma}\delta n_{\sigma} =
\delta (d^{\dagger} d-e^{\dagger} e)\,.
\end{equation}
With this at hand, the charge autocorrelation functions
can be written in terms of the slave-boson correlation functions as:
\begin{equation}
\label{korrn}
\chi_{c}(k)\!  = \sum_{\sigma\sigma^{'}}
\langle \delta n_{\sigma}(-k)\delta n_{\sigma^{'}}(k)\rangle
  = \langle \delta N (-k) \delta N(k)\rangle. 
\end{equation}
In the following calculation to one-loop order we use the notation
$k\equiv(\vec{k},\omega)$, and  the propagator $S_{ij}(k)$ as given in
Ref.~\cite{Lhou15}. The charge susceptibility results as follows:
\begin{equation}
\label{korrcc}
\chi_{c}(k) = 
2e^{2}S_{11}^{-1}(k)-4edS_{12}^{-1}(k)+2d^{2}S_{22}^{-1}(k).
\end{equation}
In the long wavelength and low frequency limit, this relation yields the Landau
parameter $F_0^s$. In contrast to the conventional random phase approximation
(RPA) results, the obtained Landau parameter cannot be brought to a simple
analytical form, unless the intersite Coulomb elements $V_{ij}$ are
neglected. In this case, simplifications are most effective at half filling
\cite{Vol84,Li91,LiBen}, where $F_0^s$ is determined by: 
\begin{equation}
\label{eqrf:f0s}
F^s_0  =  -1 + \frac{1}{(1-U/U_c)^2}\,.
\end{equation}
The impact of the Hubbard $U$ on $F_0^s$ is transparent: $F_0^s$ steadily
increases from $0$ at $U=0$ until it diverges at the metal-to-insulator
transition. It should be noted that $F^s_0$  from
Eq.~(\ref{korrcc}) is not sensitive to the details of the intersite
elements $V_{ij}$, but only to its zero-momentum Fourier transform
(Eq.~(\ref{eq:V0})). For concreteness we restrict ourselves below to nearest
neighbour interaction --- in which case $V_0 = 4 V$ --- but our results apply
as well to more general situations, provided $V$ is properly interpreted as 
$V  = \frac{1}{4} \frac{1}{L} \sum_{i,j} V_{ij}$.

\section{Results}
\label{sec:resu}

\begin{figure}[t]
\begin{minipage}{18pc}
\hskip--0.2cm\includegraphics[width=18pc]{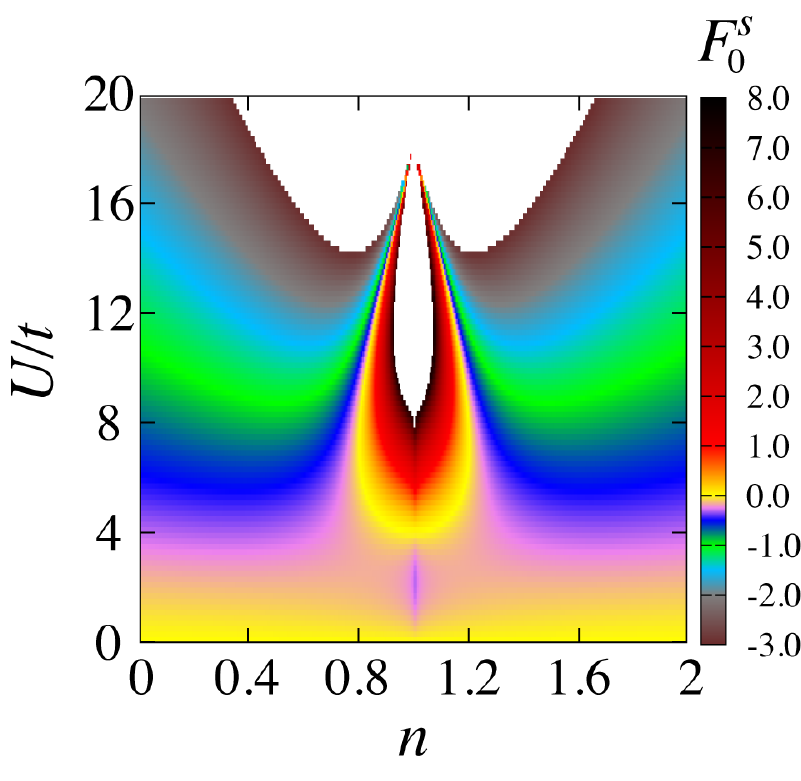}
\caption{\label{fig:f0s0mp2}Landau parameter $F_0^s$ for the square lattice as function of $n$ and
$U$. Here $t'=0$ and  $V/U = -0.2$.}
\end{minipage}\hspace{2pc}
\begin{minipage}{18pc}
\hskip-0.2cm\includegraphics[width=18pc]{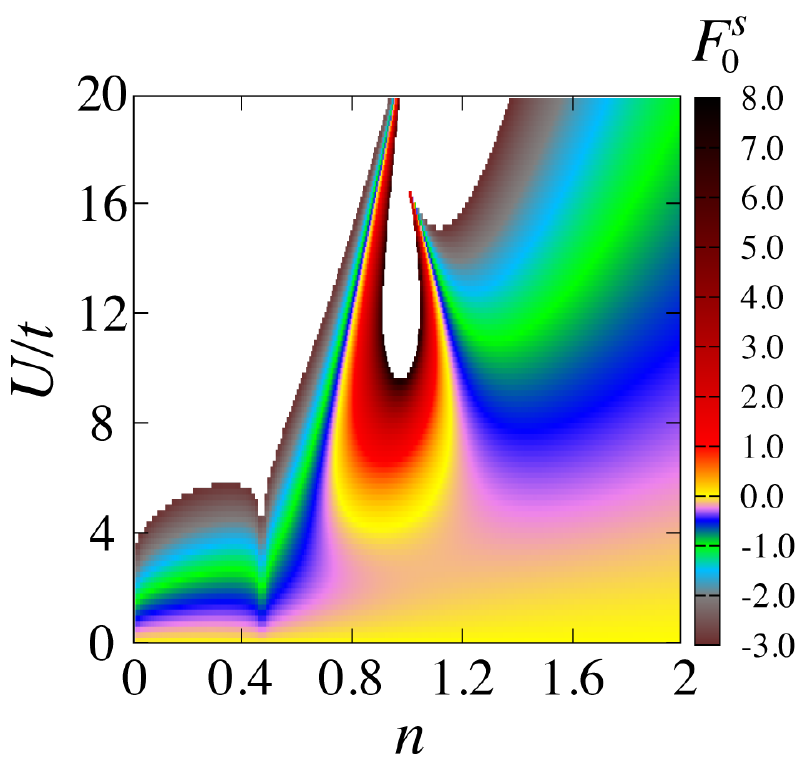}
\caption{\label{fig:f0sp15mp2}Landau parameter $F_0^s$ for the square lattice. Here $t'=-0.45\,t$ and  $V/U = -0.2$.}
\end{minipage}
\end{figure}

In this work, we evaluated the density response from the one-loop result
within slave-boson theory and extracted the dimensionless Landau parameter
$F_0^s$ in order to characterise the tendency towards a charge instability
through a single effective interaction
parameter. Our analysis builds on a 2D lattice model with two interaction
scales: A positive on-site Coulomb interaction $U$, which may be tuned to
target the  regime of strong electronic correlations, and a negative
nearest neighbour interaction $V$, which can drive a charge instability in the
case of $F_0^s \leq -1$.   

The question arises if both characteristics appear in this model
simultaneously, namely strong correlations and a charge
instability---or are they found in separate regimes that are controlled by the
electronic density $n$? Here we may choose to identify strong correlations
through a large effective mass $m^*$. In fact, $m/m^*$ is zero at half filling
and small in the respective density regime close to half filling 
(see Figs.~\ref{fig:z00} and \ref{fig:z0p45}). Moreover, $m/m^*$ is 
independent of $V$. Consistent with
Fermi liquid theory, the effective density of states at the Fermi energy
$N^*_F$ is enhanced with $m^*/m$, so that one observes three significant
structures in $N^*_F(n)$ for finite $t'$ and $U/t\gtrsim 4$ (see Fig.~\ref{fig:dos}(b)): The two lower peaks for
$n\rightarrow 0$ and $n\simeq0.5$ represent 
the Van Hove singularities of the two-dimensional DOS
for negative $t'$ (cf.~Fig.~\ref{fig:dos}), whereas the buildup of the peak at $n=1$ is a
pure correlation effect. For $t'=0$ the Van Hove peak and the correlation peak
are both placed at $n =1$, whereas a finite $t'$ serves to keep these peaks
apart (which will be helpful also in the discussion of the structure of
$F_0^s$ below). 

A graphical survey of the impact of positive $U$ and negative $V$ on
$F_0^s(n)$ is provided by Figs.~\ref{fig:f0s0mp2} and \ref{fig:f0sp15mp2}. 
We keep the ratio $V/U= -0.2$
constant in these contour plots in order to reduce the number of control
parameters. As required, $F_0^s$ is zero along the $U=0=V$ $n$-axis. For small
values of $U/t$ (and correspondingly of $-V/t$), the Landau parameter $F_0^s$
is negative but still larger than $-1$, irrespective of filling. Then, at
intermediate values of $U/t$, the Landau parameter becomes positive close to
half filling, whereas it approaches -1 for filling well below or above
half filling. The latter behaviour is induced by the negative nearest neighbour
interaction $V$, whereas the positive value of $F_0^s$ close to half filling
is a correlation effect controlled by the repulsive $U$. Eventually, at large
$U/t$ and $-V/t$, Fermi liquid behaviour breaks down, connected either with
$m/m^*=0$ at half filling or with $F_0^s\leq -1$ for finite doping away from
half filling.  

For $t'<0$  the (particle-hole) symmetry with respect to $n=1$ is broken and
the charge instability at $F_0^s<-1$ is attained already for lower values of
the interaction parameters for $n<1$ (see Fig.~\ref{fig:f0sp15mp2}). 
This enhancement of the
instability applies especially for values of $n$ at and below the density for
which the Van Hove singularity (VHs) produces a peak in the bare $N_F(n)$: A dip
structure is formed slightly below $n=0.5$ (Fig.~\ref{fig:f0sp15mp2}). 
The VHs is also reflected
in the finer dip structure in Fig.~\ref{fig:f0s0mp2} at $n=1$, however there, the 
correlation-induced increase of $F_0^s$ and the VHs-controlled decrease of 
$F_0^s$ through negative $V$ compete close to half filling. This will become evident below,
when we fix $U$ and control $F_0^s$ by $V$.   

\begin{figure}[t]
\includegraphics[width=37pc]{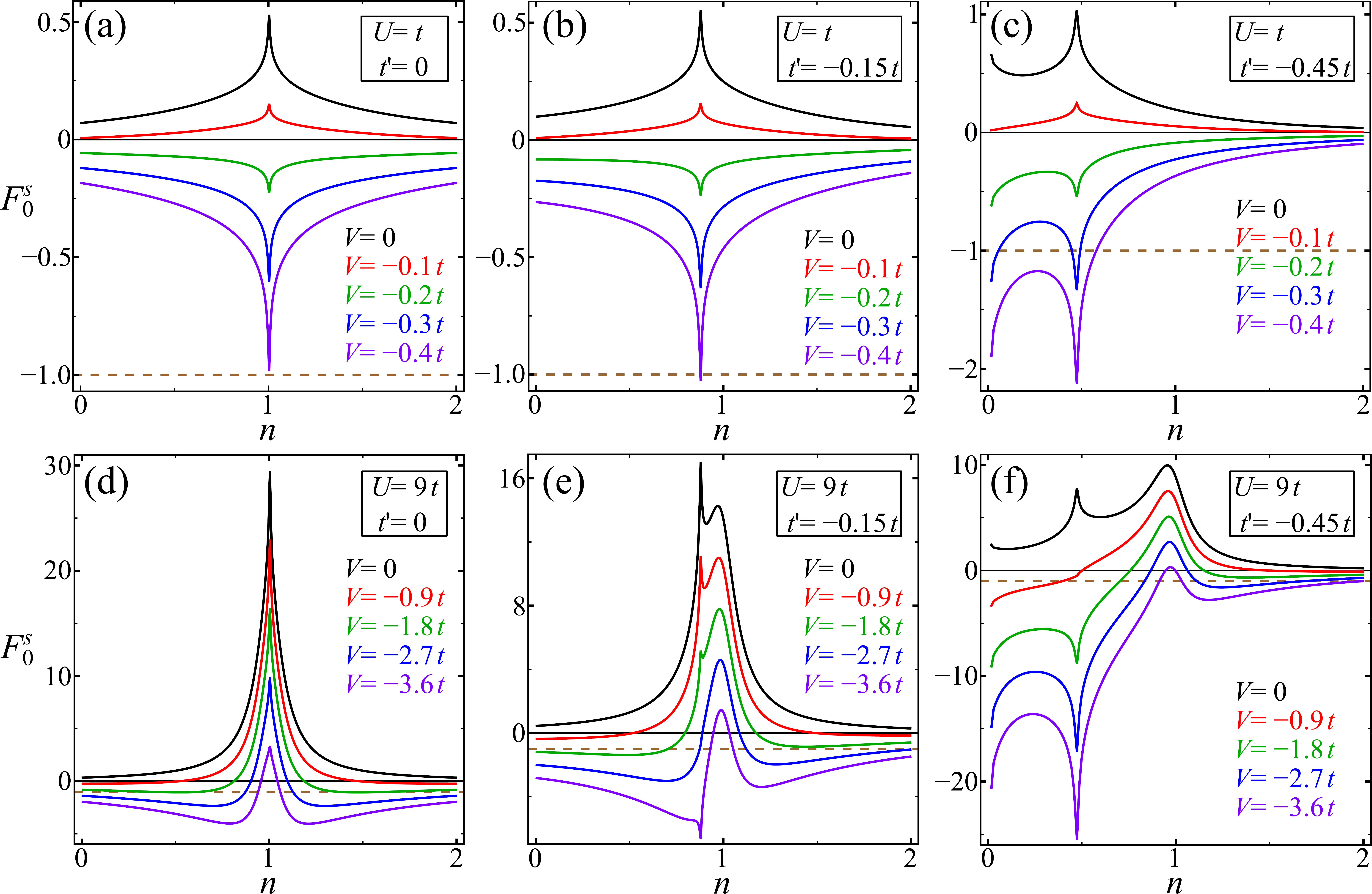}
\caption{\label{fig:F0s}Landau parameter $F_0^s$ for different sets of $t'$ and $U$ as function of filling $n$. 
The system becomes instable for $F_0^s<-1$, i.e., below the (brown) dashed line.}
\end{figure}

In Fig.~\ref{fig:F0s} we fix $U/t$ to 1 and 9 in the upper and in the lower
set of panels, respectively. In panels (a) and (d) the next-nearest neighbour hopping $t'$ is
zero. For $V/t=0$ and $V/t= -0.1$, $F_0^s$ is positive for all values of
$n$. For sufficiently negative values of $V$ ($V/t \lesssim -0.15$ for $U/t=1$
in (a)), $F_0^s$ is negative for the full range of $n$-values. 
However, for $V/t \gtrsim -0.4$ the electronic system is still stable against
charge fluctuations, as $F_0^s>-1$ still applies. 
For $U/t=9$ (panel (d)) we observe a peak around $n=1$ with $F_0^s>0$. In
contrast, $F_0^s$ is negative for $n$ not close to half filling, except for
$V=0$ when $F_0^s \geq 0$ is always true. Obviously, close to half filling
Mott-Hubbard-type electronic correlations compete with the charge instability
induced through negative $V$. It depends on the ratio of $U/|V|$ which of them
prevails. 

For $t'\neq 0$, the picture is more clear-cut (Fig.~\ref{fig:F0s}, panel (b) and (e)): For
weak coupling, that is $U/t=1$, the $n$-dependent structure of $F_0^s$ (peak
or dip) is controlled by the position of the VHs (similar to panel
(a)). However, for strong coupling $U/t=9$, a smooth peak emerges around
$n=1$. This peak was concealed in (d) by the contribution from the VHs (at
zero or small $t'$) but becomes visible when the VHs-related peak is shifted
to lower values of $n$ for more negative values of $t'$. The shape of the
function $F_0^s$ in dependence on $n$ is the central result of our work. At
half filling the Mott-Hubbard-type correlations dominate assuming that the
negative value of $V$ is not unreasonably large with respect to $U$. 

Below half filling (for negative $t'$), the position of the VHs determines the
peak or dip in $F_0^s(n)$---see also panels (c) and (f).
It is not unexpected that $F_0^s(n)$ displays a peak/dip structure controlled
by the VHs of the density of states: The dimensionless Landau parameters
are composed of a microscopic interaction times the density of states at the Fermi energy.

With this survey of $F_0^s(n)$ in the $U$-$V$ coupling parameter space,
the conclusion can be drawn that charge instabilities induced by an attractive $V$
can only be generated in a regime where strong electronic correlations are absent. 
The charge instability is boosted by VHs whereas the strong correlations effects are 
not controlled by VHs.

\section{Summary}
\label{sec:summa}

In order to gain a more thorough understanding of correlated two-dimensional
electron systems we applied slave-boson theory to an extended Hubbard model. In
the Kotliar and Ruckenstein (spin rotation invariant) slave-boson
representation it is feasible to interpolate between the non-interacting limit
and the strong coupling case. Here we used this slave-boson representation
to determine the two prominent parameters of Landau theory, $m^*/m$ and $F^s_0$. 
Beyond the one-band repulsive Hubbard model, we included a nearest neighbour
coupling $V$ which is, in our investigation, attractive as we wanted to study
instabilities in the (static) charge response. The charge response and,
correspondingly, $F^s_0$  were identified in one-loop
approximation. While we focused on the effect
  of nearest neighbour attractive interactions, it should be 
  emphasised that our results for $F^s_0$ also directly apply to more
  general situations, as the non-local interaction only enters through its
  zero-momentum Fourier component.
Furthermore, we introduced a next-nearest neighbour hopping $t'$ in order to
control the position of the Van Hove singularity (VHs)---in particular, to
shift the VHs from its position in the middle of the band (at $t'=0$) to lower
energies (for $t'<0$). Thereby we separated the correlation induced
enhancement of $F^s_0$ at half filling from DOS controlled effects. 

We found a strong enhancement of $m^*/m$ close to half filling for $U/t
\gtrsim 10$, and that mass renormalisation diverges at half filling---in agreement with the
standard approaches to the Hubbard model with strong on-site repulsion $U$.
The effective mass $m^*$ turns out to be independent of the nearest neighbour
interaction $V$, at least on the paramagnetic saddle-point level of
approximation. Moreover, $m^*/m$ depends only weakly on $t'$. As expected, the
effective mass strongly renormalises the DOS at the Fermi edge ($N_F^*$) for
half filling and intermediate to large values of $U/t$. When the VHs is
shifted towards lower energies for $t'<0$, then the peak in $N_F^*$ centred
at half filling is a signature of the correlation-induced mass enhancement,
exclusively. 

The Landau parameter $F^s_0$ is zero in the interactionless case. At any
finite $U$ and/or $V$, it is zero only for particular values of filling $n$,
in the case that repulsive $U$ and attractive $V$ compete. Higher Landau
parameters may then account for the residual interaction between
quasiparticles but this has not yet been investigated. 

For weak repulsive on-site interaction $U\simeq t$, the density dependence of
$F^s_0$ is dominated by the strong dependence on the VHs which produce peaks
of the DOS for distinct fillings. If $-V$ is larger than approximately a
tenth of $U$, the Landau parameter $F^s_0$ attains negative values in the full
range of densities. If $|V|$ is increased further, $F^s_0$ becomes less than
$-1$, primarily close to the densities where the VHs dominate, and a charge
instability emerges. 

However, for intermediate to large values of $U/t$ and $U/|V|$, the charge
instability is suppressed close to half filling. In fact, for increasing
electronic correlations, we observe the formation of a pronounced peak of
$F^s_0$ around half filling. This rounded peak with positive values is the
very signature of strong correlations and is entirely absent for weak coupling 
($U/t\simeq1$). The tendency towards charge instability for negative $V$ is
quenched by electronic correlations in this regime---below half filling it may
reemerge.  

Finally, we would like to point to the observation that the effective
interaction $F^s_0$ may still be sizable for low densities, both for
dominating repulsive $U$ or for attractive $V$. This is a hallmark of 2D
electronic systems, as in contrast $F^s_0$ vanishes in the low density limit
for generic 3D lattices, such as the cubic one
\cite{Lhou15}. This is found both in our
formalism and in perturbation theory. In that respect, it is interesting to
note that some 2D electronic systems have non-negligible interaction effects
even for low electron fillings which are not necessarily related to the
presence of long-range Coulomb interactions. Specifically, the electron system
at interfaces of LaAlO$_3$ films on SrTiO$_3$ substrates was interpreted as an
electron liquid rather than an electron gas (see
Ref.~\cite{Bre10}). Moreover the metallic state, which is formed on
SrTiO$_3$ surfaces appears to support electronic correlations, unexpected for
such a low density electron system (see Ref.~\cite{Meev11}). Further analyses
of these puzzling observations are required in this respect.

\section*{Acknowledgements}  
This work was  supported by the DFG through the TRR~80. R.F. is grateful to
the R\'egion Basse-Normandie and the Minist\`ere de la Recherche for financial
support.  
The authors acknowledge helpful discussions with V.~H.~Dao.

\end{document}